\documentclass[prl,aps,twocolumn,superscriptaddress,floatfix]{revtex4-1}
\usepackage{latexsym, amssymb, amsfonts,amsmath, amstext,graphicx, color, bbm, bm, times, hyperref, relsize, mathptm, soul,dsfont,amsthm}
\usepackage[utf8]{inputenc}
\usepackage[english]{babel}

\newcommand{\tr}{\text{tr}}

\def\tr{\mbox{tr}}
\def\bra#1{\langle{#1}|}
\def\ket#1{|{#1}\rangle}

{\catcode`\|=\active
  \gdef\Braket#1{\begingroup
\mathcode`\|32768\let|\BraVert\left<{#1}\right>\endgroup}}
\def\BraVert{\egroup\,\mid\,\bgroup}

\definecolor{kmblue}{rgb}{0.19, 0.25, 0.91}
\definecolor{kmred}{rgb}{0.79, 0.29, 0.0}
\definecolor{kmgreen}{rgb}{0, 0.42, 0.24}
\definecolor{smgreen}{rgb}{0.5, 0.22, 0.24}

\begin{document}

\title{Entropy bounds for quantum processes with initial correlations}

\author{Sai Vinjanampathy}
\email{sai@quantumlah.org}
\affiliation{Centre for Quantum Technologies, National University of Singapore, 3 Science Drive 2, Singapore 117543}

\author{Kavan Modi}
\email{kavan.modi@monash.edu}
\affiliation{School of Physics and Astronomy, Monash University, Victoria 3800, Australia}

\date{Received: \today}

\begin{abstract}
Quantum technology is progressing towards fast quantum control over systems interacting with small environments. Hence such technologies are operating in a regime where the environment remembers the system's past, and the applicability of complete-positive trace preserving maps is no longer valid. The departure from complete positivity means many useful bounds, like entropy production, Holevo, and data processing inequality are no longer applicable to such systems. We address these issues by deriving a generalized bound for entropy valid for quantum dynamics with arbitrary system-environment correlations. We employ superchannels, which map quantum operations performed by the experimenter, represented in terms of completely positive maps, to states. Our bound has information-theoretic applications, as it generalizes the data processing inequality and the Holevo bound. We prove that both data processing inequality and the Holevo are valid even when system is correlated with the environment. 
\end{abstract}

\maketitle


\section{Introduction } Physical sciences are replete with inequalities that inform us about the limits on allowed transformations. In classical equilibrium thermodynamics, the second law is an inequality that implies that the entropy produced in natural irreversible adiabatic processes is related to the heat. This second law is derived for systems in the neighborhood of thermal equilibrium, where temperature is well defined. In the absence of such simplifying assumptions, entropy production inequalities have been proposed in place of the second law. 

For completely-positive trace-preserving (CPTP) transformations, which characterize the dynamics of a system when it interacts with an independent environment, Spohn's inequality bounds the entropy generated during the process. Spohn's inequality uses relative entropy, defined as $\text{D} [\sigma_1 \Vert \sigma_2] := - \tr[ \sigma_1 \{\log(\sigma_2) - \log(\sigma_1)\}]$, and its contractivity (or monotonicity) $\text{D}[\sigma_1 \Vert\sigma_2]\geq \text{D} [\Phi(\sigma_1) \Vert \Phi(\sigma_2)]$ for any CPTP transformation $\Phi$. If $\sigma_2$ is replaced by the non-equilibrium steady state (NESS) of the CPTP transformation, i.e., $\Phi(\mathfrak{e}):=\mathfrak{e}$, the change in the von Neumann entropy $\text S(\sigma):=-\tr[\sigma \log(\sigma)]$ can be written as
\begin{align}\label{spohn}
\text{S}(\Phi(\sigma))-\text{S}(\sigma)\geq-\text{tr}[\{\Phi(\sigma)-\sigma\}\log(\mathfrak{e})].
\end{align}
The physical insight from Eq. \eqref{spohn} is that any state that is not the fixed point of the map generates a state transformation, accompanied by a change in entropy bounded by the equation above. The inequality above reduces to the standard form of the second law when the map represents an infinitesimal transformation about equilibrium, $\Phi$ is the thermal map and hence $\mathfrak{e}$ is the thermal state. However, Spohn's inequality and contractivity of relative entropy are restricted to CPTP maps only, whose construction assumes that initial state of the system is independent of the environment, i.e., the two are uncorrelated. Thus, the applicability of Eq.~\eqref{spohn} is rather limited. In this manuscript, we derive an entropy production bound without such simplifying assumptions, for generic initial system-environment state correlations. We present two important applications of the new monotonicity relationship.


\section{Dynamics beyond CPTP maps } Consider a quantum process of finite duration and the dynamics of the system between two intermediate points. This, in general, cannot be described by a CPTP map \cite{pechukas1994reduced, alicki1995comment, pechukas1995pechukas}, because the system will be correlated with its environment at any intermediate point and hence breaks the assumption that the map is independent of the state of the system. We will refer to these correlations as initial or intermediate correlations (IC). Systems with IC are non-Markovian, since IC are a record of the past interactions. Since systems with IC \cite{Weinstein2004tomography} cannot always be described by CPTP maps, a description for partial segments of a process could be very useful. For instance, imagine a qubit (the system) is reset by a cavity (the environment) after it performs some task \cite{niemczyk2010circuit}. After the first reset, the system and the environment will be correlated for a finite amount of time. These correlations may not \emph{fully} vanish by the time the qubit needs to be reset again. Such protocols are common to many quantum technologies \cite{erez2008thermodynamic, wang2011ultraefficient, wang2013absolute, PhysRevLett.112.050501}. Consequently, non-Markovian systems have drawn a lot of interest recently \cite{Rodriguez11b, breuer2009measure, chruscinski2012markovian, bylicka2013non, mazzola2012dynamical, modi2012positivity} and the presence of IC can be verified using sophisticated witnesses \cite{laine2010witness, rodriguez2012unification}, some of which have been experimentally demonstrated \cite{smirne2011experimental, li2011experimentally, PhysRevLett.114.090402}.

The presence of IC marks the departure of CPTP physics and hence the second law. Many authors have opted for giving up CP in order to deal with dynamics in presence of IC and other non-Markovian processes, leading breakdown of important physical laws. For instance, violation of the entropy production law recently demonstrated in a study of driven open quantum microcircuits \cite{argentieri2014violations}. IC also have adverse affects on quantum information theory, which is normally derived assuming no IC \cite{wilde2013quantum}. The Holevo quantity and the data processing inequality were recently shown to not hold \cite{holevo-assingment, PhysRevLett.113.140502} in this regime. IC are replete in fast quantum control experiments and in systems interacting with small environments, where it is expected that the standard rules of physics still apply. Hence the apparent violation of these aforementioned laws are an indication of a breakdown of formalism. Furthermore, given the prevalence of fast quantum control \cite{morley2013quantum}, this regime of dynamics is becoming ever more relevant for experiments. 

\begin{figure}[t]
\begin{center}
\includegraphics[width=0.30\textwidth]{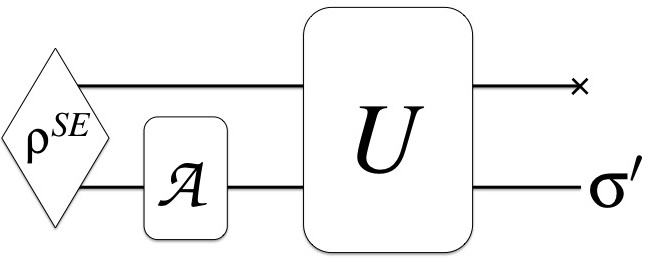}
\caption{\label{Outline} The framework of dynamics with initial or intermediate correlations (IC). An state $\rho^{SE}$ with IC is subjected to a CPTP transformation $\mathcal{A}$ which only acts on the system. The states of the environment conditioned by the measurement outcome are subjected to joint unitary dynamics along with the system state and at the final time, the ancilla state is discarded. The inequality that bounds the entropy of the final state $\sigma'$ for a choice of $\mathcal{A}$ is presented in the text.}
\end{center}
\end{figure}

The monotonicity of relative entropy, which relies on complete positivity of the dynamics, has a wide array of application:
\begin{align}
\text{CPTP} \longleftrightarrow  \text{Monotonicity} \longleftrightarrow 
\left\{\begin{matrix} \text{Entropy Production} \\ \text{Data Processing Inequality}\\ \text{Holevo Quantity} \\ \vdots \end{matrix}\right. \nonumber
\end{align}
Giving up CP means giving up a many familiar bounds like Holevo and data processing inequality. In this manuscript, using operational tool called \emph{superchannel} \cite{modi2012operational, PhysRevLett.114.090402} to describe the dynamics of a system in presence of IC~\footnote{Techniques like assignment maps \cite{pechukas1995pechukas, pechukas1994reduced, alicki1995comment, rodriguez2010linear} to address intermediate correlations suffer from non-unique representations, making them less suitable to discuss physical relevant situations such as the second law.}, we derive bounds on the entropy production bound, and restore the quantum data processing inequality and the Holevo quantity. We outline the framework of our problem in Fig.~\ref{Outline}.

\section{Operational approach to open dynamics }
We imagine a quantum process wherein the system is correlated with its environment. We denote the total state of the system and the environment as $\rho^{SE}$, and the marginal states as $\sigma$ and $\tau$ for the system and the environment respectively. An operation $\mathcal{A}$ is performed on the system, preparing the quantum state in any desired state, and conditioning the environmental state to the outcome of the system preparation. Following this, the system and environment are subjected to joint unitary evolution. The final state of the system namely $\sigma'$, is recovered by tracing over the environment at the final time. We wish to bound the entropy production of this process.

In a series of papers \cite{kuah2007state, modi2010role, modi2011preparation, modi2012operational} these dynamics were approached in an operational manner. The reasoning goes like this: Suppose we want to prepare the system into a desired pure state $\ket{\pi^{(m)}}\bra{\pi^{(m)}}$ using a specific operation, defined as $\mathcal{A}^{(m)} (\sigma) = \ket{\pi^{(m)}} \bra{\pi^{(m)}}$. Since the operation acts on the system state, the joint state after the operation is given by $\mathcal{A}^{(m)}\otimes \mathcal{I} (\rho^{SE}) =\ket{\pi^{(m)}} \bra{\pi^{(m)}} \otimes\tau^{(m)}$, where $\mathcal{I}$ is the identity map on the environment. The environmental state $\tau^{(m)}$ is conditioned on the fact that the system was prepared in state $m$. Moreover, $\tau^{(m)}$ will depend on the choice of operation $\mathcal{A}^{(m)}$. Hence, for a general quantum operation, we should care not just about the output of the preparation but the preparation procedure itself. 

After preparing the system, we let it evolve, which may include an interaction with the environment, and measure the output state $\sigma'$. The key observation in describing this dynamics is that the variable of the problem (that we are free to choose) is the operation $\mathcal{A}$, which can be \emph{any} CPTP operation. For instance, it can be a projective operation as in the example above, or a unitary operation. The corresponding output to $\mathcal{A}$ is the final state $\sigma'$. Therefore to describe the dynamics with need a map joins these two elements: $\mathcal{M}[\mathcal{A}] = \sigma'$. Map $\mathcal{M}$ is called a superchannel \cite{PhysRevLett.114.090402}, which is defined in terms of matrix indices as $\mathcal{M}_{abc;pqr}= \sum_{xyz} U_{ax;by}\; \rho^{SE}_{cy;rz}\; U^{*}_{px;qz}$, see \cite{modi2012operational} for details. It is completely positive, but not trace preserving \cite{modi2011preparation, modi2012operational}.

The superchannel's action is defined over any CPTP transformation $\mathcal{A}$. Therefore superchannels provides the most general description for the dynamics between any two time-steps \cite{modi2012operational}. Moreover, such superchannels can be fully determined via tomographic methods \cite{PhysRevLett.114.090402}. Yet, there are several ways in which superchannels differ from the familiar quantum operations relating to the fact that they transform operations to states. Firstly, superchannels are completely positive, and hence have an operator sum representation. Since quantum operations are represented by trace $d$ matrices while quantum states are unit trace operators, superchannels do not preserve trace. For every superchannel $\mathcal{M}$, this is trivially fixed by introducing another superchannel which acts on $\mathcal{A}_d=\mathcal{A}/d$, a unit trace operators, and yields the state $\sigma'$. Since the trace preserving superchannel is a composition of two CP maps, $\mathcal{M}$ and a map that transforms $\mathcal{A}$ to a state $\mathcal{A}_d$, the new superchannel is also CP, and now TP. We will refer to this modified superchannel as $\mathcal{M}^\#$.

\section{Non-Equilibrium Steady Operations } Before we present the generalization of the second law like inequality, we need a non-equilibrium steady operation (NESO) $\mathcal{E}$ that looks something like $\mathcal{M} [\mathcal{E}] =\mathcal{E}$. However, since a superchannel transforms $d^2\times d^2$ quantum operations to $d\times d$ density matrices this is not possible. In other words $\mathcal{M}^\#$ does not have a fixed point. For states without IC, the study of entropy production using quantum relative entropy compares the given state to a reference state. By extension, we seek a NESO to compare our operation against. Hence, we seek equilibria $\mathcal{E}$ such that $\mathcal{M} [\mathcal{E}] = \mathfrak{e}$, where the definition of $\mathfrak{e}$ has to be clarified. We clarify this by examining the superchannel for the familiar case of no IC, since, in this case the NESS state is well defined. Any generalized solution will have to also satisfy this special case. Here, the superchannel is written as $\mathcal{M}=\Phi \otimes \sigma$, with its action defined as $\mathcal{M} [\mathcal{A}] =\Phi (\mathcal{A} (\sigma))$. Let us then define an NESO $\mathcal{E}$ such that its action on the system state yields $\mathcal{E} (\sigma) =\mathfrak{e}$, such that this state is preserved by the rest of the dynamics, namely $\Phi(\mathfrak{e}) =\mathfrak{e}$. This step is just the condition that for a given channel and any initial state, the operation that replaces this initial state with a NESS state is a NESO. Note that since the outcome of the NESO $\mathcal{E}$ is $\mathfrak{e}$ for all initial states of the system $\sigma$, the correct NESO should be written in the uncorrelated case as $\mathcal{E}=\mathfrak{e} \otimes \mathbb{I}$. We adopt this as the NESO, with one additional condition relating to the definition of the state discussed below. From the information theoretic perspective, it is clear that $\mathcal{E}=\mathfrak{e} \otimes \mathbb I $ has the same information as $\mathfrak{e}$. Thus, the normalized superchannel acting on NESO, $\mathcal{M}^\# [\mathcal{E}_d] =\mathfrak{e}$, preserves all information.

In the case with IC, the NESO results in transforming the state of the system to an as yet undetermined state $\mathfrak{e}$, whereas it transforms the state of the environment to $\tau:=\tr_S(\rho^{SE})$. The NESS state $\mathfrak{e}$ is defined as a state that is invariant to the subsequent dynamics, namely $\mathfrak{e}:=\tr_{E}(U \mathfrak{e}\otimes\tau \, U^{\dagger})$. Physically this implies that the definition of NESO $\mathcal{E}$, is one which ``forgets" the reduced state of the system and ``replaces" it with a state that is a fixed point of the subsequent quantum channel. Just like before, the NESO $\mathcal{E}$ can be made to have unit trace by dividing through by $d$, namely $\mathcal{E}_d:=\mathcal{E}/d$. We will employ this to extend the second law.

\begin{figure}
\begin{center}
\includegraphics[width=0.40\textwidth]{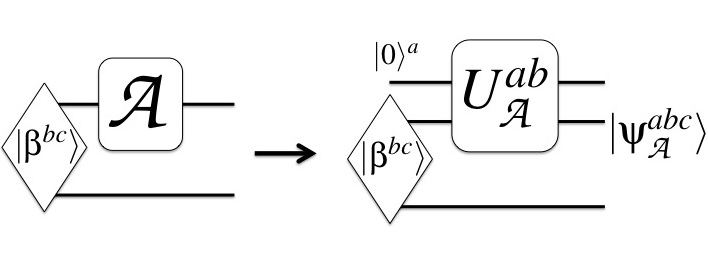}
\caption{\label{CSpring} {\bf Entropy of operations.} Consider the Choi state $\rho_\mathcal{A}$ of an operation $\mathcal{A}$. This is defined by the action of $\mathcal{A}$ on one part of a maximally entangled state $\ket{\beta^{bc}}= \sum\ket{i^bi^c}/\sqrt{d}$. This is depicted on the left. However, instead of representing the quantum operation, consider the Stinespring dilation of the operation, wherein it is represented as a joint unitary action on an ancilla: $U^{ab}_\mathcal{A} = \ket{0}^{a}\otimes\ket{\beta^{bc}}$. Since the input is a pure state and the subsequent transformation is a unitary $U^{ab}_{\mathcal{A}}$, we get at the end of this transformation a pure state $\ket{\psi^{abc}_{\mathcal{A}}}$. Now, the trace of $\ket{\psi^{abc}_{\mathcal{A}}}$ over $a$ represents the Choi state of the map $\mathcal{A}$. The reduced entropy of the subsystem $bc$ is the entropy of the map $S(\mathcal{A}_d)$, which is used in our central result in Eq. \eqref{final}. It is also the entropy of the discarded ancilla $a$, and it is a measure of the entanglement generated between the $bc$ and $a$.}
\end{center}
\end{figure}

\section{Entropy production bounds }
Since superchannels can be made CPTP, contractivity inequality can be applied to $\mathcal{M}^\#$. This results in the equation $\text{D}[\mathcal{A}_d\Vert\mathcal{E}_d]\geq\text{D}[\mathcal{M}^\#[\mathcal{A}_d]\Vert\mathcal{M}^\#[\mathcal{E}_d]]$. This can be rewritten in terms of the final state $\sigma':=\mathcal{M}^\#[\mathcal{A}_d]$ as 
\begin{align}\label{final}
\text{S}(\sigma')-\text{S}(\mathcal{A}_d)\geq
-\tr\left[\sigma'\log(\mathfrak{e})\right]+\tr\left[\mathcal{A}_d\log(\mathcal{E}_d)\right].
\end{align}
This is the the main results in this paper, which generalizes Eq.~\eqref{spohn}. The entropy of an operation is defined by entropy of its Choi state, see Fig. \ref{CSpring} caption for details. It is easy to show that Eq. \eqref{final} encompasses Eq. \eqref{spohn}. We can add the change in entropy due to the operation $\mathcal{A}$, i.e., Spohn's inequality $\text{S} (\mathcal{A} (\sigma)) -\text{S} (\sigma) \geq -\text{tr} [\{\mathcal{A} (\sigma)-\sigma\} \log(\mathfrak{a})]$, where $\mathcal{A}(\mathfrak{a}):=\mathfrak{a}$ defines the NESS of $\mathcal{A}$ to the inequality above. This inequality, alongside Eq. \eqref{final}, can be used to bound the entropy change between the reduced state of an initially correlated system and its final state. This is the first application of our main result. In the appendix, we show how to compute the entropic cost of implementing operation $\mathcal{A}$, thus fully accounting for change in system's entropy.

We now apply the bound above to derive the familiar Clausius inequality as a special case of Eq. \eqref{final}. Consider again, a generic superchannel, with the property that the channel composed of its reduced environmental state is a thermal map. Such a thermal map is characterized by its  fixed point being the thermal state $\mathfrak{e}=\exp\{-\beta(H-F)\}$, where $F=\log(Z)/\beta$ is the free energy and $Z=\tr[\exp(-\beta H)]$ is the partition function. Once again, consider the ``throw and replace" operation which replaces the correlated state of the system with an arbitrary initial state, given by $\mathcal{A}_d=\sigma\otimes\mathbb{I}_d$. Evaluating Eq. \eqref{final} with this operation, and the NESS $\mathcal{E}_d=\mathfrak{e}\otimes\mathbb{I}_d$ yields
\begin{align}
\text{S}(\sigma')-\text{S}(\sigma\otimes\mathbb{I}_d)\geq
-&\tr(\sigma'\log(\mathfrak{e}))\nonumber\\
&+\tr(\sigma\otimes\mathbb{I}_d\log(\mathfrak{e}\otimes\mathbb{I}_d)).
\end{align}
This equation, for transformations between the neighborhood of Gibbs states (so that von Neumann entropy can be associated with thermodynamic entropy and temperature is well defined), recovers the Clausius inequality. 

\section{Quantum Data Processing Inequality } Contractivity of relative entropy has two more important applications, whose generalizations we discuss as applications of our new formalism. The first is the quantum data processing inequality (QDPI), which informs us that quantum channels do not increase the mutual information between subsystems. Recently \cite{PhysRevLett.113.140502} pointed out that QDPI is violated if and only if the dynamics is not completely positive . Not-completely-positive (NCP) maps are the alternative to the superchannel formalism, however, they suffer ambiguity in terms of their operational interpretation, while the superchannel is operationally sound. Moreover, using the superchannel formalism we demonstrate that there is no need to give up complete-positivity. Here we generalize QDPI for quantum superchannels.

\begin{figure}[t]
\begin{center}
\includegraphics[width=0.45\textwidth]{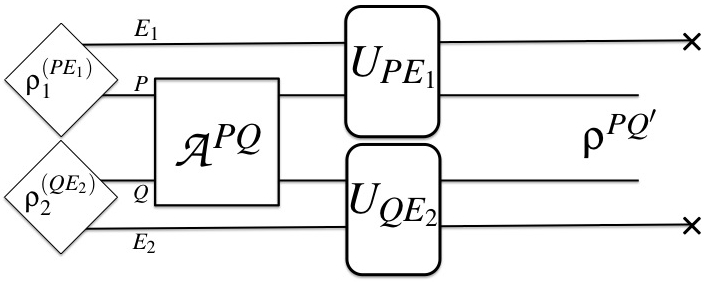}
\caption{\label{QDPI_FIG}The setting to generalize the quantum data processing inequality to states with IC. States $\rho^{(PE_1)}_1$ and $\rho^{(QE_2)}_2$ share initial correlations with their respective environments, $E_1$ and $E_2$. These states are subjected to a joint quantum operation $\mathcal{A}^{(PQ)}$. After this, the states evolve under the influence of joint unitary operations, $U_{PE_1}$ and $U_{QE_2}$ respectively. Since the initial states have IC, the dynamics is not guaranteed to be CPTP. Hence the data processing inequality seems not to apply in this case. In the text, we derive the correct form of QDPI, presented in Eq. \ref{final_qdpi}.}
\end{center}
\end{figure}

QDPI bounds the mutual information between subsystems at the output of a quantum channel to be no larger than the mutual information between the subsystems at the input. This inequality is a consequence of monotonicity. In the presence of IC, QDPI does not seem to hold. Consider the scenario in Fig. \ref{QDPI_FIG} for instance. The physical system consists of two separate quantum systems $P$ and $Q$, initially correlated with their respective environments. These systems are then subjected to a bipartite quantum operation $\mathcal{A}^{PQ}$. Following this, the individual quantum systems are allowed to interact with their respective environments, and the final bipartite state $\rho^{PQ'}$ is obtained at the end. One might seek to bound the mutual information of the final bipartite state. Intuitively we expect that this mutual information can only be related to a property of the bipartite measurement $\mathcal{A}^{PQ}$. To prove this intuition, we simply study the following form of our new second law, namely
\begin{align}
\text{D}[\mathcal{A}^{PQ}_{d^2} \Vert  & \mathcal{A}^{P}_d \otimes \mathcal{A}^{Q}_d]  \geq \\ &\text{D}\left[\mathcal{M}^{\#}_1 \otimes \mathcal{M}^{\#}_2 \left[ \mathcal{A}^{PQ}_{d^2} \right] \Vert \mathcal{M}^{\#}_1\otimes\mathcal{M}^{\#}_2 \left[ \mathcal{A}^{P}_d\otimes\mathcal{A}^{Q}_d \right]\right] \nonumber.
\end{align}
Here $\mathcal{A}^{P}$ is the bipartite measurement acting only on the subsystem $P$, and likewise $\mathcal{A}^{Q}$. $\mathcal{M}^{\#}_1$ acts on the operation that acts on system $P$ and likewise $\mathcal{M}^{\#}_2$ on $Q$. This inequality can simply be rewritten as an inequality involving mutual information, namely 
\begin{align}\label{final_qdpi}
\text{I}[P:Q]_{\rho^{PQ'}}\leq\text{I}[P:Q]_{\mathcal{A}^{PQ}_{d^2}}.
\end{align}
 The physical interpretation of this inequality is that the mutual information of the state that is output is bounded by the mutual information of the measurement performed, a simple application of our new law. This fully resolves the question as to whether initial correlations violate QDPI \cite{PhysRevLett.113.140502}.

\section{Holevo bound } The second application is related to bounding the mutual information between two communicating parties, Alice and Bob. The accessible information $I_{\text{acc}}$, which quantifies the amount of classical information that can be communicated using a quantum channel $\Phi$ is bounded by the Holevo quantity $\chi(\Phi)$. To communicate codewords $k$ with probability $p_k$, Alice sends quantum state $\sigma_k$ over a CPTP channel $\Phi$, and Bob receives $\sigma'_k = \Phi(\sigma_k)$. The Holevo quantity is then $I_{\text{acc}} \le \chi(\Phi) = \text{S}(\sum_k p_k \sigma'_k) - \sum_k p_k \text{S}(\sigma'_k).$ Recently \cite{holevo-assingment} proved that for NCP maps, the accessible information can be greater than the Holevo quantity, which is undesirable since Holevo's theorem is a central result in quantum information theory. Once again, the apparent violation of physical laws is heralding a breakdown of the formalism, which is fixed by analyzing the problem using the superchannel formalism. We present this analysis below.

Let us consider the scenario where Alice's initial state is correlated with the channel. This could be seen a backdoor attack by an adversarial party who is trying to eavesdrop. Operationally, Alice will perform operation $\mathcal{A}^{(k)}$ with probability $p_k$ on her system, and in return Bob will receive $\sigma'_k = \mathcal{M}^{\#}[\mathcal{A}^{(k)}_d]$. Since Bob receives an ensemble $\{p_k,\sigma'_k\}$, the remainder of the derivation is exactly the same as the scenario with no IC, and we recover Holevo quantity
\begin{align}
\text{I}_\text{acc}(\mathcal{M}^{\#})\leq \chi(\mathcal{M}^{\#})=&\text{S}\left(\sum \text{p}_k \mathcal{M}^{\#}[\mathcal{A}^{(k)}_d] \right) \nonumber\\
 &-\sum \text{p}_k\text{S}(\mathcal{M}^{\#}[\mathcal{A}^{(k)}_d]).
\end{align}
Once again we see that operational approach to dynamics leads to familiar results that are central to a great deal of physics and information theory. We note that in \cite{holevo-assingment}, the correlations that cause the dynamics to be not CP are shared between Alice and Bob, causing the subsequent measurements performed by Bob to be not CP. Our formalism readily accommodates this minor difference in the way the problem has been setup.
\section{Entropic cost of implementing an operation}
In this section, we discuss the cost of implementing an operation. Applying a non-unitary operation $\mathcal{A}$ requires unitarily interacting the system with an ancilla. The ancilla can be taken to be a pure state initially, and can be discarded after the interaction. However, this interaction generate correlations between the system and the ancilla, and thus discarding the ancilla has entropic costs. We will address this issue of entropy production in the ancilla by writing another second law like inequality for such dynamics. To address the ancilla, we dilate $\mathcal{A}$ to a unitary in a higher dimensional space. This is done by modifying the superchannel formalism to include the ancilla. We define an isometry as $\mathds{A} [\sigma]=V \sigma \otimes \alpha \, V^\dag$. The state of the ancilla is $\alpha$, which can be taken to be a pure state in general, and $V$ is a unitary that jointly acts on $SA$. Isometry $\mathds{A}$ is simply related to the quantum operation $\mathcal{A}$ by $\tr_A (\mathds{A}) = \mathcal{A}$. The set of isometry $\mathds{A}$ just as rich as the set of operations $\mathcal{A}$. Various choices of $V$ and $\alpha$ correspond to different operations. Just as superchannel acts on operations $\mathcal{A}$, we define $\mathds{M}$-map that acts on isometries: $\mathds{M}[\mathds{A}] = \Upsilon$, where $\Upsilon$ is the joint state of the system and ancilla. We obtain $\mathcal{M}$-map via the simple relation $\tr_A(\mathds{M} [\mathds{A}]) = \mathcal{M} [\mathcal{A}]$. The initial entropy is the entropy of the system state $\sigma$. The final entropy is the entropy of the joint state of $SA$. Then the change in entropy is $\delta S = S(\Upsilon) - S(\sigma)$. This change in entropy accounts for all resources. This formalism now, where the system and the ancilla used to perform quantum operations on the system are taken into account, can be used to write down a bound like Eq. (2).

\section{Conclusions and Outlook } We have generalized contractivity of relative entropy to include intermediate correlations of quantum systems. By doing so, we have generalized the second law like entropy inequality to such dynamics. The physical interpretation of the resulting entropy inequality was clarified with several examples and the entropy generated by the measurements used to control such correlated quantum systems was accounted for by a modified inequality. We applied the inequality to generalize the data processing inequality and the Holevo bound, two cornerstones of quantum information processing and quantum thermodynamics, to accommodate IC. The formalism of superchannels clarifies the physics of systems interacting with IC, transforming the quantum operations controlled at the input into the states output. Besides the numerous applications of the contractivity inequality, the second law like inequality, the data processing inequality and the Holevo bound, the superchannel formalism opens up systems with IC for important investigations and applications.

\begin{acknowledgments}Centre for Quantum Technologies is a Research Centre of Excellence funded by the Ministry of Education and the National Research Foundation of Singapore. We thank J. Goold for valuable discussions. This work was supported by the European COST network MP1209.
\end{acknowledgments}
%

\begin{thebibliography}{31}%
\makeatletter
\providecommand \@ifxundefined [1]{%
 \@ifx{#1\undefined}
}%
\providecommand \@ifnum [1]{%
 \ifnum #1\expandafter \@firstoftwo
 \else \expandafter \@secondoftwo
 \fi
}%
\providecommand \@ifx [1]{%
 \ifx #1\expandafter \@firstoftwo
 \else \expandafter \@secondoftwo
 \fi
}%
\providecommand \natexlab [1]{#1}%
\providecommand \enquote  [1]{``#1''}%
\providecommand \bibnamefont  [1]{#1}%
\providecommand \bibfnamefont [1]{#1}%
\providecommand \citenamefont [1]{#1}%
\providecommand \href@noop [0]{\@secondoftwo}%
\providecommand \href [0]{\begingroup \@sanitize@url \@href}%
\providecommand \@href[1]{\@@startlink{#1}\@@href}%
\providecommand \@@href[1]{\endgroup#1\@@endlink}%
\providecommand \@sanitize@url [0]{\catcode `\\12\catcode `\$12\catcode
  `\&12\catcode `\#12\catcode `\^12\catcode `\_12\catcode `\%12\relax}%
\providecommand \@@startlink[1]{}%
\providecommand \@@endlink[0]{}%
\providecommand \url  [0]{\begingroup\@sanitize@url \@url }%
\providecommand \@url [1]{\endgroup\@href {#1}{\urlprefix }}%
\providecommand \urlprefix  [0]{URL }%
\providecommand \Eprint [0]{\href }%
\providecommand \doibase [0]{http://dx.doi.org/}%
\providecommand \selectlanguage [0]{\@gobble}%
\providecommand \bibinfo  [0]{\@secondoftwo}%
\providecommand \bibfield  [0]{\@secondoftwo}%
\providecommand \translation [1]{[#1]}%
\providecommand \BibitemOpen [0]{}%
\providecommand \bibitemStop [0]{}%
\providecommand \bibitemNoStop [0]{.\EOS\space}%
\providecommand \EOS [0]{\spacefactor3000\relax}%
\providecommand \BibitemShut  [1]{\csname bibitem#1\endcsname}%
\let\auto@bib@innerbib\@empty
\bibitem [{\citenamefont {Pechukas}(1994)}]{pechukas1994reduced}%
  \BibitemOpen
  \bibfield  {author} {\bibinfo {author} {\bibfnamefont {P.}~\bibnamefont
  {Pechukas}},\ }\href@noop {} {\bibfield  {journal} {\bibinfo  {journal}
  {Phys. Rev. Lett.}\ }\textbf {\bibinfo {volume} {73}},\ \bibinfo {pages}
  {1060} (\bibinfo {year} {1994})}\BibitemShut {NoStop}%
\bibitem [{\citenamefont {Alicki}(1995)}]{alicki1995comment}%
  \BibitemOpen
  \bibfield  {author} {\bibinfo {author} {\bibfnamefont {R.}~\bibnamefont
  {Alicki}},\ }\href@noop {} {\bibfield  {journal} {\bibinfo  {journal} {Phys.
  Rev. Lett.}\ }\textbf {\bibinfo {volume} {75}},\ \bibinfo {pages} {3020}
  (\bibinfo {year} {1995})}\BibitemShut {NoStop}%
\bibitem [{\citenamefont {Pechukas}(1995)}]{pechukas1995pechukas}%
  \BibitemOpen
  \bibfield  {author} {\bibinfo {author} {\bibfnamefont {P.}~\bibnamefont
  {Pechukas}},\ }\href@noop {} {\bibfield  {journal} {\bibinfo  {journal}
  {Phys. Rev. Lett.}\ }\textbf {\bibinfo {volume} {75}},\ \bibinfo {pages}
  {3021} (\bibinfo {year} {1995})}\BibitemShut {NoStop}%
\bibitem [{\citenamefont {Weinstein}\ \emph {et~al.}(2004)\citenamefont
  {Weinstein}, \citenamefont {Havel}, \citenamefont {Emerson}, \citenamefont
  {Boulant}, \citenamefont {Saraceno}, \citenamefont {Lloyd},\ and\
  \citenamefont {Cory}}]{Weinstein2004tomography}%
  \BibitemOpen
  \bibfield  {author} {\bibinfo {author} {\bibfnamefont {Y.~S.}\ \bibnamefont
  {Weinstein}}, \bibinfo {author} {\bibfnamefont {T.~F.}\ \bibnamefont
  {Havel}}, \bibinfo {author} {\bibfnamefont {J.}~\bibnamefont {Emerson}},
  \bibinfo {author} {\bibfnamefont {N.}~\bibnamefont {Boulant}}, \bibinfo
  {author} {\bibfnamefont {M.}~\bibnamefont {Saraceno}}, \bibinfo {author}
  {\bibfnamefont {S.}~\bibnamefont {Lloyd}}, \ and\ \bibinfo {author}
  {\bibfnamefont {D.~G.}\ \bibnamefont {Cory}},\ }\href@noop {} {\bibfield
  {journal} {\bibinfo  {journal} {J. Chem. Phys.}\ }\textbf {\bibinfo {volume}
  {121}},\ \bibinfo {pages} {6117} (\bibinfo {year} {2004})}\BibitemShut
  {NoStop}%
\bibitem [{\citenamefont {Niemczyk}\ \emph {et~al.}(2010)\citenamefont
  {Niemczyk}, \citenamefont {Deppe}, \citenamefont {Huebl}, \citenamefont
  {Menzel}, \citenamefont {Hocke}, \citenamefont {Schwarz}, \citenamefont
  {Garcia-Ripoll}, \citenamefont {Zueco}, \citenamefont {H{\"u}mmer},
  \citenamefont {Solano} \emph {et~al.}}]{niemczyk2010circuit}%
  \BibitemOpen
  \bibfield  {author} {\bibinfo {author} {\bibfnamefont {T.}~\bibnamefont
  {Niemczyk}}, \bibinfo {author} {\bibfnamefont {F.}~\bibnamefont {Deppe}},
  \bibinfo {author} {\bibfnamefont {H.}~\bibnamefont {Huebl}}, \bibinfo
  {author} {\bibfnamefont {E.}~\bibnamefont {Menzel}}, \bibinfo {author}
  {\bibfnamefont {F.}~\bibnamefont {Hocke}}, \bibinfo {author} {\bibfnamefont
  {M.}~\bibnamefont {Schwarz}}, \bibinfo {author} {\bibfnamefont
  {J.}~\bibnamefont {Garcia-Ripoll}}, \bibinfo {author} {\bibfnamefont
  {D.}~\bibnamefont {Zueco}}, \bibinfo {author} {\bibfnamefont
  {T.}~\bibnamefont {H{\"u}mmer}}, \bibinfo {author} {\bibfnamefont
  {E.}~\bibnamefont {Solano}},  \emph {et~al.},\ }\href@noop {} {\bibfield
  {journal} {\bibinfo  {journal} {Nature Physics}\ }\textbf {\bibinfo {volume}
  {6}},\ \bibinfo {pages} {772} (\bibinfo {year} {2010})}\BibitemShut {NoStop}%
\bibitem [{\citenamefont {Erez}\ \emph {et~al.}(2008)\citenamefont {Erez},
  \citenamefont {Gordon}, \citenamefont {Nest},\ and\ \citenamefont
  {Kurizki}}]{erez2008thermodynamic}%
  \BibitemOpen
  \bibfield  {author} {\bibinfo {author} {\bibfnamefont {N.}~\bibnamefont
  {Erez}}, \bibinfo {author} {\bibfnamefont {G.}~\bibnamefont {Gordon}},
  \bibinfo {author} {\bibfnamefont {M.}~\bibnamefont {Nest}}, \ and\ \bibinfo
  {author} {\bibfnamefont {G.}~\bibnamefont {Kurizki}},\ }\href@noop {}
  {\bibfield  {journal} {\bibinfo  {journal} {Nature}\ }\textbf {\bibinfo
  {volume} {452}},\ \bibinfo {pages} {724} (\bibinfo {year}
  {2008})}\BibitemShut {NoStop}%
\bibitem [{\citenamefont {Wang}\ \emph {et~al.}(2011)\citenamefont {Wang},
  \citenamefont {Vinjanampathy}, \citenamefont {Strauch},\ and\ \citenamefont
  {Jacobs}}]{wang2011ultraefficient}%
  \BibitemOpen
  \bibfield  {author} {\bibinfo {author} {\bibfnamefont {X.}~\bibnamefont
  {Wang}}, \bibinfo {author} {\bibfnamefont {S.}~\bibnamefont {Vinjanampathy}},
  \bibinfo {author} {\bibfnamefont {F.~W.}\ \bibnamefont {Strauch}}, \ and\
  \bibinfo {author} {\bibfnamefont {K.}~\bibnamefont {Jacobs}},\ }\href@noop {}
  {\bibfield  {journal} {\bibinfo  {journal} {Phys. Rev. Lett.}\ }\textbf
  {\bibinfo {volume} {107}},\ \bibinfo {pages} {177204} (\bibinfo {year}
  {2011})}\BibitemShut {NoStop}%
\bibitem [{\citenamefont {Wang}\ \emph {et~al.}(2013)\citenamefont {Wang},
  \citenamefont {Vinjanampathy}, \citenamefont {Strauch},\ and\ \citenamefont
  {Jacobs}}]{wang2013absolute}%
  \BibitemOpen
  \bibfield  {author} {\bibinfo {author} {\bibfnamefont {X.}~\bibnamefont
  {Wang}}, \bibinfo {author} {\bibfnamefont {S.}~\bibnamefont {Vinjanampathy}},
  \bibinfo {author} {\bibfnamefont {F.~W.}\ \bibnamefont {Strauch}}, \ and\
  \bibinfo {author} {\bibfnamefont {K.}~\bibnamefont {Jacobs}},\ }\href@noop {}
  {\bibfield  {journal} {\bibinfo  {journal} {Phys. Rev. Lett.}\ }\textbf
  {\bibinfo {volume} {110}},\ \bibinfo {pages} {157207} (\bibinfo {year}
  {2013})}\BibitemShut {NoStop}%
\bibitem [{\citenamefont {Wood}\ \emph {et~al.}(2014)\citenamefont {Wood},
  \citenamefont {Borneman},\ and\ \citenamefont
  {Cory}}]{PhysRevLett.112.050501}%
  \BibitemOpen
  \bibfield  {author} {\bibinfo {author} {\bibfnamefont {C.~J.}\ \bibnamefont
  {Wood}}, \bibinfo {author} {\bibfnamefont {T.~W.}\ \bibnamefont {Borneman}},
  \ and\ \bibinfo {author} {\bibfnamefont {D.~G.}\ \bibnamefont {Cory}},\
  }\href {\doibase 10.1103/PhysRevLett.112.050501} {\bibfield  {journal}
  {\bibinfo  {journal} {Phys. Rev. Lett.}\ }\textbf {\bibinfo {volume} {112}},\
  \bibinfo {pages} {050501} (\bibinfo {year} {2014})}\BibitemShut {NoStop}%
\bibitem [{\citenamefont {Rodr\'iguez-Rosario}\ and\ \citenamefont
  {Sudarshan}(2011)}]{Rodriguez11b}%
  \BibitemOpen
  \bibfield  {author} {\bibinfo {author} {\bibfnamefont {C.~A.}\ \bibnamefont
  {Rodr\'iguez-Rosario}}\ and\ \bibinfo {author} {\bibfnamefont {E.~C.~G.}\
  \bibnamefont {Sudarshan}},\ }\href@noop {} {\bibfield  {journal} {\bibinfo
  {journal} {Int. J. Quant. Info.}\ }\textbf {\bibinfo {volume} {9}},\ \bibinfo
  {pages} {1617} (\bibinfo {year} {2011})},\ \Eprint
  {http://arxiv.org/abs/[arXiv:0803.1183 (2008)]} {[arXiv:0803.1183 (2008)]}
  \BibitemShut {NoStop}%
\bibitem [{\citenamefont {Breuer}\ \emph {et~al.}(2009)\citenamefont {Breuer},
  \citenamefont {Laine},\ and\ \citenamefont {Piilo}}]{breuer2009measure}%
  \BibitemOpen
  \bibfield  {author} {\bibinfo {author} {\bibfnamefont {H.-P.}\ \bibnamefont
  {Breuer}}, \bibinfo {author} {\bibfnamefont {E.-M.}\ \bibnamefont {Laine}}, \
  and\ \bibinfo {author} {\bibfnamefont {J.}~\bibnamefont {Piilo}},\
  }\href@noop {} {\bibfield  {journal} {\bibinfo  {journal} {Phys. Rev. Lett.}\
  }\textbf {\bibinfo {volume} {103}},\ \bibinfo {pages} {210401} (\bibinfo
  {year} {2009})}\BibitemShut {NoStop}%
\bibitem [{\citenamefont {Chru{\'s}ci{\'n}ski}\ and\ \citenamefont
  {Kossakowski}(2012)}]{chruscinski2012markovian}%
  \BibitemOpen
  \bibfield  {author} {\bibinfo {author} {\bibfnamefont {D.}~\bibnamefont
  {Chru{\'s}ci{\'n}ski}}\ and\ \bibinfo {author} {\bibfnamefont
  {A.}~\bibnamefont {Kossakowski}},\ }\href@noop {} {\bibfield  {journal}
  {\bibinfo  {journal} {EPL (Europhysics Letters)}\ }\textbf {\bibinfo {volume}
  {97}},\ \bibinfo {pages} {20005} (\bibinfo {year} {2012})}\BibitemShut
  {NoStop}%
\bibitem [{\citenamefont {Bylicka}\ \emph {et~al.}(2013)\citenamefont
  {Bylicka}, \citenamefont {Chru{\'s}ci{\'n}ski},\ and\ \citenamefont
  {Maniscalco}}]{bylicka2013non}%
  \BibitemOpen
  \bibfield  {author} {\bibinfo {author} {\bibfnamefont {B.}~\bibnamefont
  {Bylicka}}, \bibinfo {author} {\bibfnamefont {D.}~\bibnamefont
  {Chru{\'s}ci{\'n}ski}}, \ and\ \bibinfo {author} {\bibfnamefont
  {S.}~\bibnamefont {Maniscalco}},\ }\href@noop {} {\bibfield  {journal}
  {\bibinfo  {journal} {arXiv:1301.2585}\ } (\bibinfo {year}
  {2013})}\BibitemShut {NoStop}%
\bibitem [{\citenamefont {Mazzola}\ \emph {et~al.}(2012)\citenamefont
  {Mazzola}, \citenamefont {Rodr{\'\i}guez-Rosario}, \citenamefont {Modi},\
  and\ \citenamefont {Paternostro}}]{mazzola2012dynamical}%
  \BibitemOpen
  \bibfield  {author} {\bibinfo {author} {\bibfnamefont {L.}~\bibnamefont
  {Mazzola}}, \bibinfo {author} {\bibfnamefont {C.~A.}\ \bibnamefont
  {Rodr{\'\i}guez-Rosario}}, \bibinfo {author} {\bibfnamefont {K.}~\bibnamefont
  {Modi}}, \ and\ \bibinfo {author} {\bibfnamefont {M.}~\bibnamefont
  {Paternostro}},\ }\href@noop {} {\bibfield  {journal} {\bibinfo  {journal}
  {Phys. Rev. A}\ }\textbf {\bibinfo {volume} {86}},\ \bibinfo {pages} {010102}
  (\bibinfo {year} {2012})}\BibitemShut {NoStop}%
\bibitem [{\citenamefont {Modi}\ \emph {et~al.}(2012)\citenamefont {Modi},
  \citenamefont {Rodr{\'\i}guez-Rosario},\ and\ \citenamefont
  {Aspuru-Guzik}}]{modi2012positivity}%
  \BibitemOpen
  \bibfield  {author} {\bibinfo {author} {\bibfnamefont {K.}~\bibnamefont
  {Modi}}, \bibinfo {author} {\bibfnamefont {C.~A.}\ \bibnamefont
  {Rodr{\'\i}guez-Rosario}}, \ and\ \bibinfo {author} {\bibfnamefont
  {A.}~\bibnamefont {Aspuru-Guzik}},\ }\href@noop {} {\bibfield  {journal}
  {\bibinfo  {journal} {Phys. Rev. A}\ }\textbf {\bibinfo {volume} {86}},\
  \bibinfo {pages} {064102} (\bibinfo {year} {2012})}\BibitemShut {NoStop}%
\bibitem [{\citenamefont {Laine}\ \emph {et~al.}(2010)\citenamefont {Laine},
  \citenamefont {Piilo},\ and\ \citenamefont {Breuer}}]{laine2010witness}%
  \BibitemOpen
  \bibfield  {author} {\bibinfo {author} {\bibfnamefont {E.-M.}\ \bibnamefont
  {Laine}}, \bibinfo {author} {\bibfnamefont {J.}~\bibnamefont {Piilo}}, \ and\
  \bibinfo {author} {\bibfnamefont {H.-P.}\ \bibnamefont {Breuer}},\
  }\href@noop {} {\bibfield  {journal} {\bibinfo  {journal} {EPL (Europhysics
  Letters)}\ }\textbf {\bibinfo {volume} {92}},\ \bibinfo {pages} {60010}
  (\bibinfo {year} {2010})}\BibitemShut {NoStop}%
\bibitem [{\citenamefont {Rodr{\'\i}guez-Rosario}\ \emph
  {et~al.}(2012)\citenamefont {Rodr{\'\i}guez-Rosario}, \citenamefont {Modi},
  \citenamefont {Mazzola},\ and\ \citenamefont
  {Aspuru-Guzik}}]{rodriguez2012unification}%
  \BibitemOpen
  \bibfield  {author} {\bibinfo {author} {\bibfnamefont {C.~A.}\ \bibnamefont
  {Rodr{\'\i}guez-Rosario}}, \bibinfo {author} {\bibfnamefont {K.}~\bibnamefont
  {Modi}}, \bibinfo {author} {\bibfnamefont {L.}~\bibnamefont {Mazzola}}, \
  and\ \bibinfo {author} {\bibfnamefont {A.}~\bibnamefont {Aspuru-Guzik}},\
  }\href@noop {} {\bibfield  {journal} {\bibinfo  {journal} {EPL (Europhysics
  Letters)}\ }\textbf {\bibinfo {volume} {99}},\ \bibinfo {pages} {20010}
  (\bibinfo {year} {2012})}\BibitemShut {NoStop}%
\bibitem [{\citenamefont {Smirne}\ \emph {et~al.}(2011)\citenamefont {Smirne},
  \citenamefont {Brivio}, \citenamefont {Cialdi}, \citenamefont {Vacchini},\
  and\ \citenamefont {Paris}}]{smirne2011experimental}%
  \BibitemOpen
  \bibfield  {author} {\bibinfo {author} {\bibfnamefont {A.}~\bibnamefont
  {Smirne}}, \bibinfo {author} {\bibfnamefont {D.}~\bibnamefont {Brivio}},
  \bibinfo {author} {\bibfnamefont {S.}~\bibnamefont {Cialdi}}, \bibinfo
  {author} {\bibfnamefont {B.}~\bibnamefont {Vacchini}}, \ and\ \bibinfo
  {author} {\bibfnamefont {M.~G.}\ \bibnamefont {Paris}},\ }\href@noop {}
  {\bibfield  {journal} {\bibinfo  {journal} {Phys. Rev. A}\ }\textbf {\bibinfo
  {volume} {84}},\ \bibinfo {pages} {032112} (\bibinfo {year}
  {2011})}\BibitemShut {NoStop}%
\bibitem [{\citenamefont {Li}\ \emph {et~al.}(2011)\citenamefont {Li},
  \citenamefont {Tang}, \citenamefont {Li},\ and\ \citenamefont
  {Guo}}]{li2011experimentally}%
  \BibitemOpen
  \bibfield  {author} {\bibinfo {author} {\bibfnamefont {C.-F.}\ \bibnamefont
  {Li}}, \bibinfo {author} {\bibfnamefont {J.-S.}\ \bibnamefont {Tang}},
  \bibinfo {author} {\bibfnamefont {Y.-L.}\ \bibnamefont {Li}}, \ and\ \bibinfo
  {author} {\bibfnamefont {G.-C.}\ \bibnamefont {Guo}},\ }\href@noop {}
  {\bibfield  {journal} {\bibinfo  {journal} {Phys. Rev. A}\ }\textbf {\bibinfo
  {volume} {83}},\ \bibinfo {pages} {064102} (\bibinfo {year}
  {2011})}\BibitemShut {NoStop}%
\bibitem [{\citenamefont {Ringbauer}\ \emph {et~al.}(2015)\citenamefont
  {Ringbauer}, \citenamefont {Wood}, \citenamefont {Modi}, \citenamefont
  {Gilchrist}, \citenamefont {White},\ and\ \citenamefont
  {Fedrizzi}}]{PhysRevLett.114.090402}%
  \BibitemOpen
  \bibfield  {author} {\bibinfo {author} {\bibfnamefont {M.}~\bibnamefont
  {Ringbauer}}, \bibinfo {author} {\bibfnamefont {C.~J.}\ \bibnamefont {Wood}},
  \bibinfo {author} {\bibfnamefont {K.}~\bibnamefont {Modi}}, \bibinfo {author}
  {\bibfnamefont {A.}~\bibnamefont {Gilchrist}}, \bibinfo {author}
  {\bibfnamefont {A.~G.}\ \bibnamefont {White}}, \ and\ \bibinfo {author}
  {\bibfnamefont {A.}~\bibnamefont {Fedrizzi}},\ }\href {\doibase
  10.1103/PhysRevLett.114.090402} {\bibfield  {journal} {\bibinfo  {journal}
  {Phys. Rev. Lett.}\ }\textbf {\bibinfo {volume} {114}},\ \bibinfo {pages}
  {090402} (\bibinfo {year} {2015})}\BibitemShut {NoStop}%
\bibitem [{\citenamefont {Argentieri}\ \emph {et~al.}(2014)\citenamefont
  {Argentieri}, \citenamefont {Benatti}, \citenamefont {Floreanini},\ and\
  \citenamefont {Pezzutto}}]{argentieri2014violations}%
  \BibitemOpen
  \bibfield  {author} {\bibinfo {author} {\bibfnamefont {G.}~\bibnamefont
  {Argentieri}}, \bibinfo {author} {\bibfnamefont {F.}~\bibnamefont {Benatti}},
  \bibinfo {author} {\bibfnamefont {R.}~\bibnamefont {Floreanini}}, \ and\
  \bibinfo {author} {\bibfnamefont {M.}~\bibnamefont {Pezzutto}},\ }\href@noop
  {} {\bibfield  {journal} {\bibinfo  {journal} {EPL}\ }\textbf {\bibinfo
  {volume} {107}},\ \bibinfo {pages} {50007} (\bibinfo {year}
  {2014})}\BibitemShut {NoStop}%
\bibitem [{\citenamefont {Wilde}(2013)}]{wilde2013quantum}%
  \BibitemOpen
  \bibfield  {author} {\bibinfo {author} {\bibfnamefont {M.~M.}\ \bibnamefont
  {Wilde}},\ }\href@noop {} {\emph {\bibinfo {title} {Quantum information
  theory}}}\ (\bibinfo  {publisher} {Cambridge University Press},\ \bibinfo
  {year} {2013})\BibitemShut {NoStop}%
\bibitem [{\citenamefont {Masillo}\ \emph {et~al.}(2011)\citenamefont
  {Masillo}, \citenamefont {Scolarici},\ and\ \citenamefont
  {Solombrino}}]{holevo-assingment}%
  \BibitemOpen
  \bibfield  {author} {\bibinfo {author} {\bibfnamefont {F.}~\bibnamefont
  {Masillo}}, \bibinfo {author} {\bibfnamefont {G.}~\bibnamefont {Scolarici}},
  \ and\ \bibinfo {author} {\bibfnamefont {L.}~\bibnamefont {Solombrino}},\
  }\href {\doibase http://dx.doi.org/10.1063/1.3525832} {\bibfield  {journal}
  {\bibinfo  {journal} {Journal of Mathematical Physics}\ }\textbf {\bibinfo
  {volume} {52}},\ \bibinfo {eid} {012101} (\bibinfo {year}
  {2011})}\BibitemShut {NoStop}%
\bibitem [{\citenamefont {Buscemi}(2014)}]{PhysRevLett.113.140502}%
  \BibitemOpen
  \bibfield  {author} {\bibinfo {author} {\bibfnamefont {F.}~\bibnamefont
  {Buscemi}},\ }\href {\doibase 10.1103/PhysRevLett.113.140502} {\bibfield
  {journal} {\bibinfo  {journal} {Phys. Rev. Lett.}\ }\textbf {\bibinfo
  {volume} {113}},\ \bibinfo {pages} {140502} (\bibinfo {year}
  {2014})}\BibitemShut {NoStop}%
\bibitem [{\citenamefont {Morley}\ \emph {et~al.}(2013)\citenamefont {Morley},
  \citenamefont {Lueders}, \citenamefont {Mohammady}, \citenamefont {Balian},
  \citenamefont {Aeppli}, \citenamefont {Kay}, \citenamefont {Witzel},
  \citenamefont {Jeschke},\ and\ \citenamefont {Monteiro}}]{morley2013quantum}%
  \BibitemOpen
  \bibfield  {author} {\bibinfo {author} {\bibfnamefont {G.~W.}\ \bibnamefont
  {Morley}}, \bibinfo {author} {\bibfnamefont {P.}~\bibnamefont {Lueders}},
  \bibinfo {author} {\bibfnamefont {M.~H.}\ \bibnamefont {Mohammady}}, \bibinfo
  {author} {\bibfnamefont {S.~J.}\ \bibnamefont {Balian}}, \bibinfo {author}
  {\bibfnamefont {G.}~\bibnamefont {Aeppli}}, \bibinfo {author} {\bibfnamefont
  {C.~W.}\ \bibnamefont {Kay}}, \bibinfo {author} {\bibfnamefont {W.~M.}\
  \bibnamefont {Witzel}}, \bibinfo {author} {\bibfnamefont {G.}~\bibnamefont
  {Jeschke}}, \ and\ \bibinfo {author} {\bibfnamefont {T.~S.}\ \bibnamefont
  {Monteiro}},\ }\href@noop {} {\bibfield  {journal} {\bibinfo  {journal}
  {Nature materials}\ }\textbf {\bibinfo {volume} {12}},\ \bibinfo {pages}
  {103} (\bibinfo {year} {2013})}\BibitemShut {NoStop}%
\bibitem [{\citenamefont {Modi}(2012)}]{modi2012operational}%
  \BibitemOpen
  \bibfield  {author} {\bibinfo {author} {\bibfnamefont {K.}~\bibnamefont
  {Modi}},\ }\href@noop {} {\bibfield  {journal} {\bibinfo  {journal}
  {Scientific reports}\ }\textbf {\bibinfo {volume} {2}},\ \bibinfo {pages}
  {581} (\bibinfo {year} {2012})}\BibitemShut {NoStop}%
\bibitem [{Note1()}]{Note1}%
  \BibitemOpen
  \bibinfo {note} {Techniques like assignment maps \cite {pechukas1995pechukas,
  pechukas1994reduced, alicki1995comment, rodriguez2010linear} to address
  intermediate correlations suffer from non-unique representations, making them
  less suitable to discuss physical relevant situations such as the second
  law.}\BibitemShut {Stop}%
\bibitem [{\citenamefont {Kuah}\ \emph {et~al.}(2007)\citenamefont {Kuah},
  \citenamefont {Modi}, \citenamefont {Rodriguez-Rosario},\ and\ \citenamefont
  {Sudarshan}}]{kuah2007state}%
  \BibitemOpen
  \bibfield  {author} {\bibinfo {author} {\bibfnamefont {A.-M.}\ \bibnamefont
  {Kuah}}, \bibinfo {author} {\bibfnamefont {K.}~\bibnamefont {Modi}}, \bibinfo
  {author} {\bibfnamefont {C.~A.}\ \bibnamefont {Rodriguez-Rosario}}, \ and\
  \bibinfo {author} {\bibfnamefont {E.}~\bibnamefont {Sudarshan}},\ }\href@noop
  {} {\bibfield  {journal} {\bibinfo  {journal} {Phys. Rev. A}\ }\textbf
  {\bibinfo {volume} {76}},\ \bibinfo {pages} {042113} (\bibinfo {year}
  {2007})}\BibitemShut {NoStop}%
\bibitem [{\citenamefont {Modi}\ and\ \citenamefont
  {Sudarshan}(2010)}]{modi2010role}%
  \BibitemOpen
  \bibfield  {author} {\bibinfo {author} {\bibfnamefont {K.}~\bibnamefont
  {Modi}}\ and\ \bibinfo {author} {\bibfnamefont {E.~C.~G.}\ \bibnamefont
  {Sudarshan}},\ }\href@noop {} {\bibfield  {journal} {\bibinfo  {journal}
  {Phys. Rev. A}\ }\textbf {\bibinfo {volume} {81}},\ \bibinfo {pages} {052119}
  (\bibinfo {year} {2010})}\BibitemShut {NoStop}%
\bibitem [{\citenamefont {Modi}(2011)}]{modi2011preparation}%
  \BibitemOpen
  \bibfield  {author} {\bibinfo {author} {\bibfnamefont {K.}~\bibnamefont
  {Modi}},\ }\href@noop {} {\bibfield  {journal} {\bibinfo  {journal} {Open
  Systems \& Information Dynamics}\ }\textbf {\bibinfo {volume} {18}},\
  \bibinfo {pages} {253} (\bibinfo {year} {2011})}\BibitemShut {NoStop}%
\bibitem [{\citenamefont {Rodr{\'\i}guez-Rosario}\ \emph
  {et~al.}(2010)\citenamefont {Rodr{\'\i}guez-Rosario}, \citenamefont {Modi},\
  and\ \citenamefont {Aspuru-Guzik}}]{rodriguez2010linear}%
  \BibitemOpen
  \bibfield  {author} {\bibinfo {author} {\bibfnamefont {C.~A.}\ \bibnamefont
  {Rodr{\'\i}guez-Rosario}}, \bibinfo {author} {\bibfnamefont {K.}~\bibnamefont
  {Modi}}, \ and\ \bibinfo {author} {\bibfnamefont {A.}~\bibnamefont
  {Aspuru-Guzik}},\ }\href@noop {} {\bibfield  {journal} {\bibinfo  {journal}
  {Phys. Rev. A}\ }\textbf {\bibinfo {volume} {81}},\ \bibinfo {pages} {012313}
  (\bibinfo {year} {2010})}\BibitemShut {NoStop}%
\end{thebibliography}
%

\end{document}